\documentclass[%
twocolumn,amsmath,amssymb
reprint,
superscriptaddress,
amsmath,amssymb,
pra,
]{revtex4}

\usepackage{graphicx}
\usepackage{dcolumn}
\usepackage{bm}
\usepackage{float}
\usepackage{threeparttable}
\usepackage{upquote}  
\usepackage{multirow}

\def\beq{\begin{equation}}
\def\eeq{\end{equation}}
\def\bea{\begin{eqnarray}}
\def\eea{\end{eqnarray}}
\def\brcl{\begin{array}{rcl}}
\def\bccl{\begin{array}{ccl}}
\def\blcl{\begin{array}{lcl}}
\def\err{\end{array}}

\begin{document}

\preprint{APS/123-QED}

\title{Operator quantum machine learning: \\ 
Navigating the chemical space of response properties}

\author{Anders S. Christensen}
\affiliation{Institute of Physical Chemistry and National Center for Computational Design and Discovery of Novel Materials (MARVEL), Department of Chemistry, University of Basel, 4056 Basel, Switzerland}
\author{O. Anatole von Lilienfeld}
\email{anatole.vonlilienfeld@unibas.ch}
\affiliation{Institute of Physical Chemistry and National Center for Computational Design and Discovery of Novel Materials (MARVEL), Department of Chemistry, University of Basel, 4056 Basel, Switzerland}

\date{\today}

\begin{abstract}
The identification and use of structure property relationships lies at the heart of the chemical sciences.
Quantum mechanics forms the basis for the unbiased virtual exploration of chemical compound space (CCS), imposing substantial compute needs if chemical accuracy is to be reached. 
In order to accelerate predictions of quantum properties without compromising accuracy, 
our lab has been developing quantum machine learning (QML) based models which can be applied throughout CCS.
Here, we briefly explain, review, and discuss the recently introduced operator formalism which 
substantially improves the data efficiency for QML models of common response properties.
\end{abstract}

\pacs{Valid PACS appear here}
\maketitle


\section{Introduction}
Chemical compound space (CCS) comprises all the theoretically possible molecules and materials one can conceive of. 
Just for medium sized organic drug like molecules it was estimated to exceed 10$^{60}$~\cite{ChemicalSpace}, 
more than atoms in our solar system~\cite{mullard2017drug}. 
Corresponding thermodynamic and kinetic stability, as well as most other observable properties can be calculated, often to 
a fair degree of accuracy, using well established approximations and frameworks of quantum and statistical mechanics~\cite{JensenCompChem,tuckerman_book_SM}. 
Unfortunately, however, the resulting numerical complexity is so dramatic that a first principles 
based exploration of representative swaths of CCS remains prohibitive for all but its smallest subsets. 
This computational burden can be reduced through the use of machine learning (ML) which, after regressing 
on a sufficiently large training data set, affords models with controlled test errors and milli-second prediction speed. 

While ML has had a long tradition in the chem- and bio-informatics communities for a long time, 
its application to potential energy surface fitting for the purpose of vibrational analysis,
reaction dynamics, or molecular dynamics already started in the nineties using 
neural networks (NN)~\cite{SumpterNoidNeuralNetworks1992} and kernel methods~\cite{Rabitz1996}. 
By contrast, inferring solutions of the electronic Schr\"odinger equation throughout compound space 
was only made possible using Kernel Ridge Regression (KRR) and neural networks (NN) in 2012~\cite{CM} 
and 2013~\cite{Montavon2013}, respectively. 
Corresponding demonstrations for crystals followed shortly after~\cite{schutt2014represent,MLcrystals_Felix2015,Elpasolite_2016,schmidt2017predicting},
and ever since, a substantial number of papers on this topic has been published in this rapidly growing field. 
For more details, we refer interested readers to the recent machine learning issues in the {\em Int. J. Quantum Chem.}~\cite{RuppForeword2015}
and {\em J. Chem. Phys.}~\cite{rupp2018guest}, as well as the overviews given in
Refs.~\cite{anatole-ijqc2013,fpdesign2014anatole,RaghusReview2016,huang2018quantum,QMLessayAnatole,faber2019modeling}.
Due to the rigorous link to the underlying physical laws of quantum and statistical mechanics, we have dubbed this approach 
``Quantum Machine Learning'' (QML), implying that we aim to model quantum properties using classical ML algorithms. 
This usage of terms is in complete analogy to Quantum Monte Carlo (using classical Monte Carlo to sample electron configuration space), 
or Quantum Molecular Dynamics (integrating Newton's equations of motion with quantum chemically computed forces).
Note, however, that this is not to be confused with ML algorithms implemented on quantum computers.

\section{Learning curves}
For any QML model to be functional, it must {\em learn}, i.e.~its predictive power {must} improve with increasing training set size $N$. 
Vapnik and others showed decades ago that ML models must, if set-up properly, exhibit this behaviour and converge down to { arbitrary} accuracy in the limit of infinite training set size~\cite{vapnik2013book}.
In practice, however, arbitrary accuracy is neither required, nor do we operate in large data set limits. Therefore the relevant question for any distribution of data points consists of asking how much more predictive power is to be gained for each additional training instance.
This dependence is manifested in learning curves which report prediction (or test) error $E$ as a function of training set size,
as discussed for KRR~\cite{vapnik1994learningcurves}, as well as NNs~\cite{StatError_Muller1996}.

In accordance with standard ML protocol~\cite{AssessmentMLJCTC2013},  
we stress the importance of (i) prediction errors being obtained exclusively on hold-out data, i.e.~for input and output samples which have never been seen during training, and (ii) hyper-parameters, noise-levels, and overfitting being accounted for through cross-validation { prior} to prediction error evaluation.
Leading prediction error terms are known to be inversely proportional to training set size, $E \propto a/N^b$, implying convenient linear behavior on 
log-log scales~\cite{vapnik1994learningcurves}.  
Off-set $\log(a)$ and slope $b$ enable the assessment and comparison of different QML models, and imply that 
prediction errors of newly developed QML models have to be reported for at least three training set sizes in order to be 
(a) meaningful, and (b) demonstrate linearity.
Absence of linearity on log-log curves typically indicates severe problems and can be due to at least one of any of the following short-comings:
\begin{itemize}
\item[(i)] Not all input variables are being accounted for (e.g. due to overly coarse representations), resulting in lack of uniqueness. 
\item[(ii)] The model is underdetermined due to lack of model complexity (too few parameters, too rigid functional form). 
\item[(iii)] Noisy or inconsistent data. 
\end{itemize} 

Meeting the uniqueness criterion is a necessary condition for meaningful models in chemistry (proof is given in Ref.~\cite{FourierDescriptor}). 
This is is not always obvious, however, since it also depends on subtleties in the nature of the target property which defines
which degrees of freedom are being sampled.  
For example, molecular graph based representations, such as SMILES strings, can be appropriate for generating
QML models of free energies of solvation for those temperature and pressure combinations for which
all conformational degrees of freedom are being averaged out while all covalent bonds are being conserved.
However, when it comes to properties which are functions of instantaneous geometries (clamped nuclei), 
e.g.~HOMO/LUMO eigenvalues or atomic forces, SMILES are too coarse, and representations encoding {all} degrees of
freedom in a unique fashion are necessary.

Low QML model off-sets $\log(a)$ and steep slopes $b$ are highly desirable in order to maximize data efficiency and thereby
minimize training data needs in order to maximize transferability through CCS. 
A considerable amount of recent work has been devoted to this endeavour, demonstrating how to lower off-sets through
ever improving representations~\cite{BOB,BAML,HDAD,pronobis2018many,FCHL}, the use of 
hierarchies among more approximate base line quantum chemistry models~\cite{DeltaPaper2015,Zaspel2018boosting},
or optimal training data-selection~\cite{Nick2016GA}.
Steeper slopes have also been obtained through use AMons (training on local building-blocks selected on the fly)~\cite{Amons}, 
as well as through the operator QML approach discussed below.

\section{Operator approach}
In 2015, we demonstrated that using the same representation and kernel, KRR based QML models can account for {\em any} QM property~\cite{SingleKernel2015}.
By analogy with basic quantum mechanics (knowledge about a single wavefunction enables evaluation of expectation values for any property's operator)
this finding suggest that the kernel assumes the role of the wavefunction, while KRR regression coefficients carry the units of the property in question.
These results represent a meaningful baseline with all QML models being functional for all properties 
(all learning curves indicated constant and negative slopes $b$). 
We noticed, however, that slope and off-set were more favorable for some properties than for others. 
In 2017, similar observations were also made for other representations and regressors, including KRR, NNs, and random forests~\cite{HDAD}.

%
It is known that learning curves improve upon explicit encoding of invariant relationships between input features and labels.
For example, it is crucial to use rotational invariant features in order to predict the energy of molecules accurately~\cite{Montavon_NIPS2012}.
Motivated by the apparent discrepancy of QML models for atomization energies of small organic molecules reaching  predictive 
power after training on only thousands of examples, and the relatively worse performance of the same models for predicting 
other properties,\cite{HDAD} we developed and introduced the Operator QML (OQML) framework~\cite{OQML}.
Within kernel-based regression,\cite{MueMikRaeTsuSch01,scholkopf2002learning,Vovk2013,HasTibFri01}
we decompose the total potential energy $U^*$ of a query molecule in its electronic ground-state, 
into a sum of atomic contributions $E_I$ for each atom $I$,
\begin{align}
        U^{*} = \sum_{I \in C} E_I\left(q^{*}_I\right) 
        =\sum_{I \in i} \sum_{J} \alpha_J k(q_J, q^{*}_I) \label{eq:local_decomposition}
\end{align}
where $q$ is an atomic environment in some chosen representation basis, $J$ is training atom $J$, and $\alpha_J$ is its regression weight.
As a side note, a comparative discussion of energies of atoms in molecules based on QML, basis-set overlap, 
and alchemical perturbation DFT has just recently been published~\cite{AtomicAPDFT}.

The OQML approach extends this KRR model while still ensuring that regression coefficients are also obtained in closed-form.
Within this framework, the relationship between a property, typically the energy, and any of its derivatives up 
to any order (typical response properties) is explicitly enforces through the application of \textit{response operators} to the learning formalism.
More specifically and using matrix notation, the  response property $\omega$, i.e.~an observable which can be computed by applying a differential operator $\mathcal{O}$ acting 
on the energy $U^{*}$, can be approximated by a finite kernel as
\begin{align}
{\mathbf{\omega}} = \mathcal{O} [\mathbf{U}] \approx \mathcal{O}[\mathbf{K}]\bm{\alpha}.
\end{align}
Here, we assume that the regression coefficients do not depend on the perturbation
(in complete analogy to force-field parameters, or the electron density for first order perturbation theory within quantum mechanics).

Then, the set of regression coefficients, $\bm{\alpha}$ minimizes the Lagrangian
\begin{widetext}
\begin{align}\label{eq:lstsq2}
J(\bm{\alpha}) & = & \sum_{\gamma} \beta_{\gamma} \| \mathcal{O}_{\gamma}[\mathbf{U}^{\rm ref}] - \mathcal{O}_{\gamma}[\mathbf{K} \bm{\alpha} ] \|^2_{L_2(\Omega_{\gamma})} \; \;
\equiv \;\; \sum_{\gamma} \beta_{\gamma} \int_{\Omega_{\gamma}} \Big[\mathcal{O}_{\gamma}[\mathbf{U}^{\rm ref}] - \mathcal{O}_{\gamma}[\mathbf{K} \bm{\alpha} ]\Big]^T\Big[\mathcal{O}_{\gamma}[\mathbf{U}^{\rm ref}] - \mathcal{O}_{\gamma}[\mathbf{K} \bm{\alpha} ] \Big] 
\end{align}
\end{widetext}
over some training set of known derivative values of $\mathcal{O}[\mathbf{U}^{\rm ref}]$. 
$\Omega_{\gamma}$ is the domain over which the corresponding operator should be minimized, e.g.~all rotational degrees of freedom if the operator acts on a SO(3) group. $\gamma$ denotes the specific perturbation, so that the model can be trained for multiple properties simultaneously, for example energies, gradients, and dipole moments.
Assuming only one perturbation, the resulting expression for the regression coefficients $\bm{\alpha}$ which solves the associated normal equations reads, 
\begin{align}
\label{eq:lsqr_alpha}
\bm{\alpha}  & = & 
\left( \int_{\Omega} \mathcal{O}[\mathbf{K}]^T  \mathcal{O}[\mathbf{K}] \right)^{-1}  \left(\int_{\Omega} \mathcal{O}[\mathbf{U}^{\rm ref}]^T\mathcal{O}[\mathbf{K}]  \right) 
\end{align}
In comparison to the expression for regression coefficients within conventional KRR, it is clear from these equations that kernel derivatives must 
play a crucial role in the loss function and thereby constraining the resulting regression solution towards a more meaningful manifold of solutions from which the most probable is being obtained. 
For further details and discussions, we refer the reader to the original work~\cite{OQML,FCHL19}.

Maybe the most frequent application concerned deals with derivatives with respect to atomic displacement, i.e. forces.
Much improved learning curves for forces are obtained when compared to KRR based QML force models 
which were trained directly on labels corresponding to each force vector elements after 
rotating each atom into a common reference frame with unique handiness, 
as done for atoms displaced along normal modes for the first time in 2015~\cite{MLatoms_2015}.
Within the most recently revision of FCHL, FCHL19~\cite{FCHL19}, we have provided a
detailed analysis of the OQML based force model which can be applied throughout CCS. 
Using the correct operators to generate appropriate loss-functions ensures that important physics is automatically being enforced, 
such as rotational covariance of forces, and that the resulting force models conserve energy. 
A comparative discussion of other state of the art KRR force models, such as 
SOAP~\cite{BartokGabor_Descriptors2013,CeriottiScienceUnified2017} and GDML~\cite{chmiela2017machine,chmiela2019sgdml} 
has also been included in Ref.~\cite{FCHL19}. 

\begin{figure}
    \centering
    \includegraphics[width=0.8\linewidth]{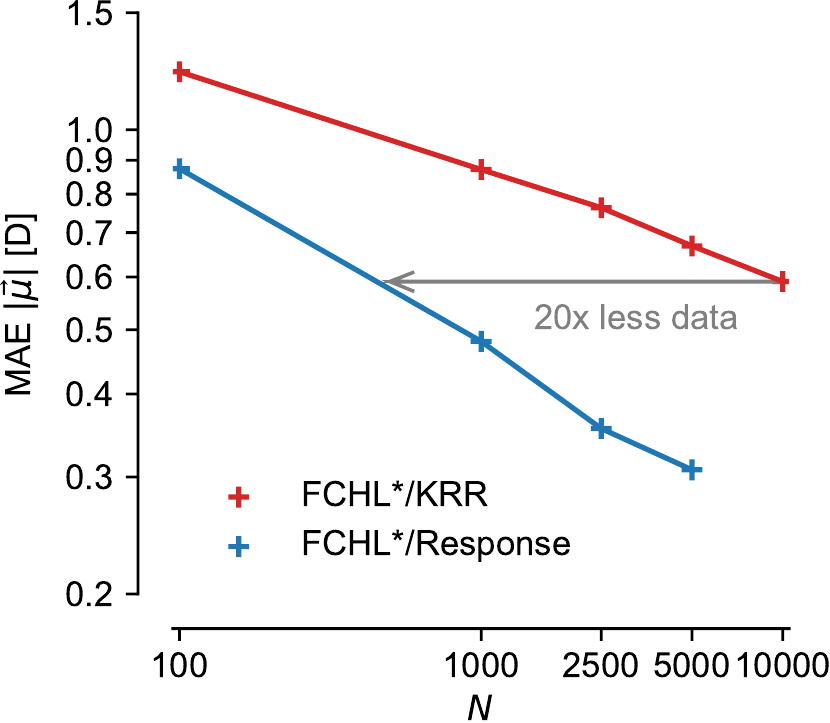}
    \caption{\label{fig:dipole_learning} 
Improvement of learning curves of dipole moment predictions in QM9 data set~\cite{QM9} with (blue) and without (red) response formalism. 
Both QML models use the FCHL$^{*}$  representation and a Gaussian kernel function, 
and differ only by their loss function and the resulting choice of regressor, 
i.e. KRR (red) and OQML (blue). 
Figure reproduced from Ref.~\cite{OQML}, licensed under a Creative Commons Attribution (CC BY) license.
        }
\end{figure}

Learning curves on display in Fig.\ref{fig:dipole_learning} exemplify 
the improvements gained when exploiting the differential relationships through OQML.
Learning curves are shown for QML models of the dipole norm of small organic molecules with up to nine atoms of the elemental types CNOF from the QM9 dataset, 
saturated with hydrogen atoms~\cite{QM9}.
The corresponding response operator to the dipole is the negative derivative with respect to a change in an external electric field.
Results for two kernel based models are shown, both using the FCHL representation\cite{FCHL,OQML} 
for the molecules and a Gaussian kernel function.
When training the conventional KRR model, the dipole norm is the only scalar label and the loss function merely regresses its deviation from reference. 
By contrast in the case of OQML, the operator is applied to the loss function of the energy property before the regression (requiring the extension
of the representation to also include a simplified electric field model, denoted as FCHL$^{*}$). 
This results in a loss function with very different terms which trivially reproduces the orientational dependence of the 
applied field and which yields QML models with much improved off-set and slope of learning curve:  
The OQML-based machine learns the dipole norm to the same accuracy as KRR with 20 times less data.

Similarly, in Ref.~\cite{OQML} we have also demonstrated how OQML can 
be applied to the second order derivatives defining Hessians which, together with dipole-moments, 
has enabled the first QML based direct prediction of an infrared spectrum from vibrational analysis.

\section{Conclusions}
We have briefly motivated the use of QML models, and highlighted the role of learning curves. 
Thereafter, the OQML approach was described, and illustrated for the case of dipole moment predictions. 
We stress that, in analogy to perturbation theory going beyond the Hellmann-Feynman theorem, 
OQML is sufficiently general to also go beyond first order derivatives. 
In fact, it is straight-forward to apply the OQML-approach to any property which is a derivative of the energy (or any other reference property), 
as long as the representation of the molecule can be perturbed accordingly.
If and how this framework can also be used for operators other than differential operators remains to be studied.

\section*{Acknowledgement}
We acknowledge support by the Swiss National Science foundation (No.~PP00P2\_138932, 407540\_167186 NFP 75 Big Data, 200021\_175747), and from the European Research Council (ERC-CoG grant QML). This work was partly supported by the NCCR MARVEL, funded by the Swiss National Science Foundation. 

\bibliography{literatur}{}
\bibliographystyle{ieeetr}

\end{document}